
\magnification \magstep1
\raggedbottom
\openup 4\jot
\voffset6truemm
\headline={\ifnum\pageno=1\hfill\else
\hfill{\it Spontaneously broken SU(5) symmetries and one-loop
effects in the early universe}
\hfill \fi}
\rightline {DSF-91/19}
\centerline {\bf SPONTANEOUSLY BROKEN SU(5) SYMMETRIES}
\centerline {\bf AND ONE-LOOP EFFECTS}
\centerline {\bf IN THE EARLY UNIVERSE$^{*}$}
\vskip 2cm
\centerline {FRANCO BUCCELLA$^{1}$, GIAMPIERO
ESPOSITO$^{2,1}$ and GENNARO MIELE$^{1,2}$}
\centerline {${ }^{(1)}$Dipartimento di Scienze Fisiche,
Mostra d'Oltremare Padiglione 19, 80125 Napoli, Italy}
\centerline {${ }^{(2)}$Istituto Nazionale di Fisica Nucleare,
Gruppo IV Sezione di Napoli,}
\centerline {Mostra d'Oltremare Padiglione 20, 80125 Napoli, Italy}
\vskip 1cm
\centerline {\bf October 1991}
\vskip 1cm
\noindent
{\bf Abstract.} This paper studies one-loop effective potential and
spontaneous-symmetry-breaking pattern for $SU(5)$ gauge
theory in De Sitter space-time. Curvature effects modify
the flat-space effective potential by means of a very
complicated special function previously derived in the
literature. An algebraic technique already developed by
the first author to study spontaneous symmetry breaking
of $SU(n)$ for renormalizable polynomial potentials
is here generalized, for $SU(5)$, to the much harder case
of a De Sitter background. A detailed algebraic and
numerical analysis provides a better derivation of the stability of
the extrema in the maximal subgroups $SU(4)\times U(1)$,
$SU(3)\times SU(2)\times U(1)$,
$SU(3)\times U(1)\times U(1) \times R_{311}$,
$SU(2)\times SU(2)\times U(1)\times U(1) \times R_{2211}$, where
$R_{311}$ and $R_{2211}$ discrete symmetries select particular
directions in the corresponding two-dimensional strata.
One thus obtains a deeper
understanding of the result, previously found with a different numerical
analysis, predicting the slide of the
inflationary universe into either the $SU(3)\times SU(2)\times U(1)$
or $SU(4)\times U(1)$ extremum. Interestingly, using this approach,
one can easily generalize all previous results to a more complete
$SU(5)$ tree-level potential also containing cubic terms.
\vskip 14cm
\noindent
$^{*}$ Classical and Quantum Gravity, {\bf 9}, 1499-1509 (1992).
\vskip 100cm
\leftline {\bf 1. Introduction}
\vskip 1cm
In the cosmological standard model [1], one assumes that
gravity is described by Einstein's general relativity,
and that the observed universe is spatially homogeneous
and isotropic. Moreover, if the energy-momentum tensor
takes a perfect-fluid form, Einstein's equations lead in
particular to the following differential equation
governing the time evolution of the cosmic scale factor
$a(t)$ :
$$
{\Bigr({\dot a \over a}\Bigr)}^{2}+{k\over a^{2}}=
{8 \pi \over 3}G \rho \; \; \; \; ,
\eqno (1.1)
$$
where $k=+1,0,-1$ respectively for a closed, flat or
open universe, $G$ is Newton's constant, and $\rho$ is
the energy density. In the matter-dominated era $\rho$
is proportional to $a^{-3}$, and in the radiation-dominated
era $\rho$ is proportional to $a^{-4}$.

The model here outlined, however, leads to a paradox :
the universe would contain about $10^{84}$ regions causally
disconnected, although its large-scale properties are
described by the Friedmann-Robertson-Walker geometry.
Moreover, denoting by $\rho_{cr}$ the energy-density value
separating an open from a closed universe, one would find
$$
{{\mid \rho - \rho_{cr} \mid} \over \rho} < 10^{-55}
\; \; \; \; .
\eqno (1.2)
$$
This is a severe fine-tuning problem, since condition
(1.2) does not seem to arise by virtue of general principles,
and appears as an {\it ad hoc} extra assumption.

However, as shown in [2], one might hope to solve these
problems (cf. [3,4]) if the cosmic scale factor $a(t)$
grows exponentially in the early universe, rather than
following the $t^{\gamma}$-behaviour of the cosmological
standard model. This can be achieved if the right-hand side
of Eq. (1.1) is constant, since this implies
$$
a(t)=a_{0}\exp \Bigr(\sqrt{{8\pi \over 3}G\rho_{0}} \; t\Bigr)
\; \; \; \; , \; \; \; \;
t \in ]t_{0},t_{a}[
\; \; \; \; ,
\eqno (1.3)
$$
provided the effect of ${k\over a^{2}}$ can be neglected
in the interval $]t_{0},t_{a}[$.
One can then show that causally disconnected
regions would no longer occur (although
severe inhomogeneities can be shown to remain [5]).
For this purpose, we need at least
a (massive [5,6], or massless self-interacting [7]) scalar
field, or a more complete theory of matter fields providing
a large vacuum-energy density $(>>M_{W})$ which drives {\it inflation},
i.e. the evolution of $a(t)$ described by Eq. (1.3).
If Eq. (1.3) holds, the corresponding
geometry is the one of De Sitter space-time, the Lorentzian
four-manifold with $R \times S^{3}$ topology and constant positive
scalar curvature.

The naturally occurring candidates for a very fundamental theory
which provides at the same time the unification of electro-weak and
strong interactions, and a suitably large vacuum energy (see above)
for symmetry-breaking are the GUTs [5,8].
Although the minimal $SU(5)$ theory [9] has been ruled
out by proton-decay experiments [10,11], the study of this $SU(5)$ model
may be very instructive. Moreover, it is
worth bearing in mind  that $SU(5)$ is contained in $SO(10)$ and $E_{6}$ [8].

We here study the one-loop effective potential
to determine the phase to which the early universe
eventually evolves [12,13]. Since we are interested in
quantum-field-theory calculations, we use the Wick-rotated
path-integral approach, and we work on the real,
Riemannian section of the corresponding complex space-time
manifold. Note that this rotation does not affect the
effective potential, while making the perturbative theory
well-defined (see below).
We are thus interested in the Riemannian version of the
De Sitter manifold, with $S^{4}$ topology. Its metric is
smooth and positive-definite, and the action of the
non-abelian Yang-Mills-Higgs theory here studied involves
elliptic, self-adjoint, positive-definite differential
operators leading to Gaussian integrals,
so that the corresponding {\it one-loop}
calculations are well-defined, even though the {\it full}
quantum theory via path integrals does not seem to have
rigorous mathematical foundations. Note that we are not
quantizing gravity, but we study quantized matter fields
in a fixed, curved, Riemannian background geometry via
Wick-rotated path integrals and perturbation theory.

Our paper is thus organized as follows. Sect. 2 describes the minimal
$SU(5)$ model in De Sitter space and the corresponding results
for the one-loop effective potential [13]. Sect. 3 presents
the basic results about the tree-level Higgs potential for
$SU(5)$ gauge theory in flat space [14]. The special
function ${\cal A}$ occurring in the corresponding one-loop
calculation in a De Sitter background is then studied in
detail. Sect. 4 provides the generalization of the technique
used in [14] to a De Sitter background.
Absolute minima are derived using both analytic
and numerical calculations, improving the understanding obtained
in [13]. Exact, approximate and asymptotic formulae for
the one-loop effective potential are shown to shed
new light on the $SU(5)$ symmetry-breaking pattern. Finally,
the concluding remarks are presented in Sect. 5.
\vskip 10cm
\leftline {\bf 2. $SU(5)$ model in De Sitter space}
\vskip 1cm
Following the introduction and [13], the bare Lagrangian ${\cal L}_{0}$
and the renormalizable tree potential of our $SU(5)$ Yang-Mills-Higgs theory
in De Sitter space are taken to be respectively
(after analytic continuation to the Riemannian manifold
with $S^{4}$ topology)
$$
{\cal L}_{0}={1\over 4}Tr \Bigr({\bf F}_{\mu \nu}{\bf F}^{\mu \nu}\Bigr)
+{1\over 2}Tr \left[ \Bigr(D_{\mu} {\bf \Phi} \Bigr)
\Bigr(D^{\mu}{\bf \Phi} \Bigr)^{\dagger} \right]+V_{0}({\bf \Phi})
\; \; \; \; ,
\eqno (2.1)
$$
$$
V_{0}({\bf \Phi})={\xi \over 2}R \;
Tr \Bigr({{\bf \Phi}}^{2}\Bigr)
+\Lambda_{2}{\Bigr(Tr {{\bf \Phi}}^{2}\Bigr)}^{2}
+\Lambda_{4}\Bigr(Tr{{\bf \Phi}}^{4}\Bigr)
\; \; \; \; ,
\eqno (2.2)
$$
where ${\bf F}_{\mu \nu} \equiv \nabla_{\mu}{\bf A}_{\nu}
-\nabla_{\nu}{\bf A}_{\mu}-ig \Bigr[{\bf A}_{\mu},{\bf A}_{\nu}\Bigr]$,
and $D_{\mu}{\bf \Phi} \equiv \partial_{\mu} {\bf \Phi}
-ig \Bigr[{\bf A}_{\mu},{\bf \Phi} \Bigr]$. Note that the
covariant derivative $\nabla_{\mu}$ differs from $\partial_{\mu}$ for terms
involving Christoffel symbols [1], and $V_{0}({\bf \Phi})$ is assumed
to obey the symmetry $V_{0}({\bf \Phi}) = V_{0}(-{\bf \Phi})$.
Moreover, as usual, $g$ is the
dimensionless coupling constant and $R={12\over r^{2}}$ is the scalar
curvature of De Sitter space ($r$ being the four-sphere radius).

The Higgs scalar field ${\bf \Phi}$ is
assumed to be in the adjoint representation of $SU(5)$ [13].
The presence in the minimal $SU(5)$ model
of an additional representation $({\bf {\underline 5}})$ of scalar fields
${\bf H}$, necessary to break the symmetry down to
$SU(3)_{C} \times U(1)_{Q}$, is irrelevant for the inflationary
scheme, due to the smaller mass value $M_{H} \approx M_{W}$.

The background-field method is now applied
to obtain the one-loop form of the potential, writing the
Higgs field as ${\bf \Phi}_{0}+{\widetilde {\bf \Phi}}$, where
${\bf \Phi}_{0}$ is a constant background field and
${\widetilde {\bf \Phi}}$ a fluctuation around
${\bf \Phi}_{0}$ (and similarly for ${\bf A}^{\mu}$). As explained
in [13], it is convenient to choose t'Hooft's gauge-fixing
term
$$
{\cal L}_{G.F.}={\alpha \over 2}
Tr {\Bigr(\nabla_{\mu}{\bf A}^{\mu}-ig {\alpha}^{-1}
[{\bf \Phi}_{0},{\bf \Phi}]\Bigr)}^{2}
\; \; \; \; ,
\eqno (2.3)
$$
and Coleman-Weinberg's theory can be used to neglect the
contribution of all scalar-field loop diagrams. This implies
that only gauge-field loop diagrams are relevant.
A very convenient form of the one-loop potential is
obtained using the gauge invariance of the theory which
enables one to diagonalize the scalar field ${\bf \Phi}$.
The corresponding diagonal form of ${\bf \Phi}$ is here denoted
by $\hat {\bf \Phi}=diag \Bigr(\varphi_{1},\varphi_{2},\varphi_{3},
\varphi_{4},\varphi_{5}\Bigr)$, where
$\sum_{i=1}^{5}{\varphi_{i}}=0$.
Thus, denoting by $\psi(t)$ the special function
${\Gamma'(t)\over \Gamma(t)}$, and defining
$$ \eqalignno{
{\cal A}(z) & \equiv {z^{2}\over 4}+{z\over 3}
-\int_{2}^{{3\over 2}+\sqrt{{1\over 4}-z}}
t \Bigr(t-{3\over 2}\Bigr)(t-3)\psi(t) \; dt \cr
& -\int_{1}^{{3\over 2}-\sqrt{{1\over 4}-z}}
t \Bigr(t-{3\over 2}\Bigr)(t-3)\psi(t) \; dt
\; \; \; \; ,
&(2.4)\cr}
$$
the one-loop effective potential for the minimal
$SU(5)$ model is found to be [13]
$$ \eqalignno{
V({\hat {\bf \Phi}})&={15\over 64 \pi^{2}}
\left \{Q+{1\over 3} \Bigr(1-\log(r^{2}M_{X}^{2})\Bigr)
\right \} R \; g^{2} {\parallel {\hat {\bf \Phi}} \parallel}\cr
&+\left \{ {9\over 128 \pi^{2}}
\Bigr(1-\log(r^{2}M_{X}^{2})\Bigr)
-{21\over 320 \pi^{2}}\Lambda \right \}
g^{4}{\parallel {\hat {\bf \Phi}} \parallel}^{2}\cr
&+{15\over 128 \pi^{2}}
\left \{ {12\over 5}\Lambda +
\Bigr(1-\log(r^{2}M_{X}^{2})\Bigr) \right \}
g^{4}\sum_{i=1}^{5}{\varphi_{i}}^{4}\cr
&-{3\over 16\pi^{2}r^{4}}\sum_{i,j=1}^{5}
{\cal A}\left[{r^{2}g^{2}\over 2}
{(\varphi_{i}-\varphi_{j})}^{2} \right] \; \; \; \; ,
&(2.5)\cr}
$$
where [13]
$$
Q \equiv {32 \pi^{2} \over 15 g^{2}}
\left [\xi-{8\over 5g^{2}}
\Bigr(\Lambda_{2}+{7\over 30}\Lambda_{4}\Bigr)\right]
\; \; \; \; ,
\eqno (2.6)
$$
$$
\Lambda \equiv {64\pi^{2}\over 15 g^{4}}
\Bigr({3\over 5}\Lambda_{4}-\Lambda_{2}\Bigr)
\; \; \; \; ,
\eqno (2.7)
$$
$$
\parallel {\bf {\Phi}} \parallel \equiv \sum_{i=1}^{5}
{\varphi}_{i}^{2}
\; \; \; \; ,
\eqno (2.8)
$$
and $M_{X}$ is related to the dimensional parameter $\mu$
appearing in the (regularized) one-loop amplitudes [12].
Moreover, if $\xi={1\over 6}$, the Higgs field is
conformally coupled to gravity.

The one-loop potential $V({\hat {\bf \Phi}})$ is then used to
determine broken-symmetry phases and curved-space phase
diagrams as shown in [13]. As a result of his numerical
analysis, the author of [13] found what follows :

(1) In the $SU(5)$ theory, the universe, in addition to the right
$SU(3)\times SU(2)\times U(1)$ direction, is also likely to
end up in the wrong $SU(4)\times U(1)$ phase;

(2) The $SU(2)\times SU(2)\times U(1)\times U(1) \times R_{2211}$ and
$SU(3)\times U(1)\times U(1) \times R_{311}$ phases are unstable for any
values of the parameters appearing in the model.

As we said in the introduction, the aim of this paper is
to provide a better
understanding of the results obtained in [13]. For this
purpose, we recall some basic results about
spontaneous symmetry breaking of $SU(n)$ [14], and about the
${\cal A}$ function [13] defined in Eq. (2.4).
\vskip 10cm
\leftline {\bf 3. Polynomial potentials and the ${\cal A}$ function}
\vskip 1cm
To study the spontaneous-symmetry-breaking
directions of the potential in Eq. (2.5), it is convenient
to define the variables
$$
a_{i} \equiv {g \; r\over \sqrt{2}} \varphi_{i}
\; \; \; \; .
\eqno (3.1)
$$
For our purpose, it is not strictly needed to study the part
of the potential depending on the norm of the
${\bf a}$ field :
${\parallel {\bf a} \parallel} \equiv \sum_{i=1}^{5}a_{i}^{2}$.
The relevant part of the potential is instead given by (up
to the multiplicative constant
${3r^{-4}\over 16 \pi^{2}}$)
$$
V_{M} \equiv b \sum_{i=1}^{5}a_{i}^{4}-
\sum_{i,j=1}^{5}{\cal A}\Bigr[(a_{i}-a_{j})^{2}\Bigr]
\; \; \; \; ,
\eqno (3.2)
$$
$$
b \equiv 6\Lambda +{5\over 2}
\Bigr(1-\log(r^{2}M_{X}^{2})\Bigr)
\; \; \; \; .
\eqno (3.3)
$$
As a first step, it is useful to recall the exact results [14]
holding for a theory where $V_{M}$ is only given by the first
term on the right-hand side of Eq. (3.2). In that case, since
$\sum_{i=1}^{5}a_{i}$ is set to zero, and $\sum_{i=1}^{5}a_{i}^{2}$
equals ${\parallel {\bf a} \parallel}$ by definition,
the Lagrange-multipliers technique can be used to
study the third-order algebraic equations leading to the
calculation of the minima [14].

The corresponding results yield, for the minima with the
residual symmetry :
$$
{\bf a}=a_{321} \equiv {{\parallel {\bf a} \parallel}^{1\over 2}
\over \sqrt{30}} \;
diag (2,2,2,-3,-3) \; \; \; \;   , \; \; \; \;
(SU(3) \times SU(2) \times U(1)) \; \; \; \; ,
\eqno (3.4)
$$
$$
{\bf a}=a_{2211} \equiv {{\parallel {\bf a} \parallel}^{1\over 2}  \over 2}
\; diag (1,1,0,-1,-1) \; \; , \; \;
(SU(2) \times SU(2) \times U(1) \times U(1) \times R_{2211}) \; \; ,
\eqno (3.5)
$$
$$
{\bf a}=a_{311} \equiv {{\parallel {\bf a} \parallel}^{1\over 2} \over
\sqrt{2}} \; diag (0,0,0,1,-1) \; \; , \; \;
(SU(3) \times U(1) \times U(1) \times R_{311}) \; \; ,
\eqno (3.6)
$$
$$
{\bf a}=a_{41} \equiv {{\parallel {\bf a} \parallel}^{1\over 2} \over
\sqrt{20}} \; diag (1,1,1,1,-4) \; \; \; \; , \; \; \; \;
(SU(4) \times U(1)) \; \; \; \; ,
\eqno (3.7)
$$
the hierarchy
$$
V_{M}(a_{41})>V_{M}(a_{311})>V_{M}(a_{2211})>V_{M}(a_{321})
\eqno (3.8)
$$
if $b>0$, and the reversed inequalities if $b<0$. Thus, when
the Higgs field is in the adjoint representation, the $SU(5)$
symmetry breaking leads only to the $SU(4)\times U(1)$ or
$SU(3)\times SU(2) \times U(1)$ symmetric minima.

Since the complete $V_{M}$ potential is in our case given by
Eqs. (3.2-3), we need to study in detail the contribution of the
${\cal A}$ function. While performing this analysis, it is useful to
supplement definition (2.4) by the Taylor expansion of
${\cal A}$ as $z \rightarrow 0$, and its asymptotic expansion as
$z \rightarrow \infty$, which are given respectively by [13]
$$ \eqalignno{
{\cal A}(z) & = 2\left(\gamma -{1\over 3}\right)z +{(\gamma -1)\over 2}z^{2}
+{z^{3}\over 6} \Bigr(-5+4\zeta(3)\Bigr)\cr
&+{z^{4}\over 24}\Bigr(-36+30\zeta(3)\Bigr)
+O(z^{5}) \; \; \; \; ,
&(3.9)\cr}
$$
$$
{\cal A}(z) \sim - \left({z^{2}\over 4}+z + {19\over 30}\right)\log(z)
+{3\over 8}z^{2} + z
\; \; \; \; ,
\eqno (3.10)
$$
where $\gamma$ is Euler's constant and
$\zeta$ is the Riemann zeta-function [12].
Using Eqs. (2.4) and (3.9-10), we have found inequalities
analogous to (3.8). In other words, defining
$$
{\cal A}(a_{41}) \equiv -\sum_{i,j=1}^{5}
{\cal A}\Bigr[(a_{i}-a_{j})^{2}\Bigr]_{{\bf a}=a_{41}}
\; \; \; \; ,
\eqno (3.11)
$$
and similarly for the other phases, one finds
$$
{\cal A}(a_{41})>{\cal A}(a_{311})>{\cal A}(a_{2211})>
{\cal A}(a_{321})
\; \; \; \; ,
\eqno (3.12)
$$
where
$$ \eqalignno{
{\cal A}(a_{41})
&=-8{\cal A}\left({5\over 4}{\parallel {\bf a} \parallel}\right)
\; \; \; \; , \cr
&{\cal A}(a_{311})=-12{\cal A}\left({{\parallel {\bf a} \parallel}
\over 2}\right)
-2{\cal A}\Bigr(2{\parallel {\bf a} \parallel}\Bigr) \; \; \; \; , \cr
&{\cal A}(a_{2211})=-8{\cal A}\left({\parallel {\bf a} \parallel}
\over 4\right)
-8{\cal A}\Bigr({\parallel {\bf a} \parallel}\Bigr) \; \; \; \; , \cr
&{\cal A}(a_{321})=-12{\cal A}\left({5\over 6}
{\parallel {\bf a} \parallel}\right) \; \; \; \; .
&(3.13,14,15,16)\cr}
$$
The inequalities appearing in Eq. (3.12) are illustrated in
Figures 1-3.
\vskip 1cm
\leftline {\bf 4. Absolute minima}
\vskip 1cm
For fixed values of the bare parameters $\xi, \Lambda_{2},
\Lambda_{4}$ and $M_{X}$ (cf. Eqs. (2.6,7) and (3.3)),
$b$ depends on $r$ as shown in Eq. (3.3). Thus
in the early universe, at small values
of $r$, i.e. when the scalar curvature is very large, $b$ is
positive, whereas it may become negative
as $r$ increases.

As shown in Sect. 3, when $b>0$, the two terms of the $V_{M}$
potential in Eq. (3.2) follow the inequalities (3.8) and (3.12).
This implies that in the very early universe the only possible phase
transition is $SU(5) \rightarrow SU(3) \times SU(2) \times U(1)$.

By contrast, for suitably large values of $r$, $b$ becomes negative,
and the polynomial part of the $V_{M}$ potential is then dominant.
In this case the analysis in [14] holds, and the phase transition
occurs in the $SU(4) \times U(1)$ direction (i.e. the previous
hierarchy is inverted).

A more detailed analysis is however in order when $b<0$ but
$\mid b \mid$ is not too large. For this purpose, using the
Taylor expansion (3.9) up to third-order, we begin by studying
the range of validity of the inequalities
$$
V_{M}(a_{41})>V_{M}(a_{311})>V_{M}(a_{2211})>V_{M}(a_{321})
\; \; \; \; .
\eqno (4.1)
$$
Thus, defining $\Omega \equiv {(-5+4\zeta(3))\over 6} <0$, one
finds
$$
\Bigr[V_{M}(a_{41})-V_{M}(a_{321})\Bigr]>0
\Longleftrightarrow
12b+60(1-\gamma)>-250 \mid \Omega \mid \;
{\parallel {\bf a} \parallel}  \; \; \; \; ,
\eqno (4.2)
$$
$$
\Bigr[V_{M}(a_{311})-V_{M}(a_{321})\Bigr]>0
\Longleftrightarrow
12b+60(1-\gamma)>-475 \mid \Omega \mid \;
{\parallel {\bf a} \parallel} \; \; \; \; ,
\eqno (4.3)
$$
$$
\Bigr[V_{M}(a_{2211})-V_{M}(a_{321})\Bigr]>0
\Longleftrightarrow
12b+60(1-\gamma)>-850 \mid \Omega \mid \;
{\parallel {\bf a} \parallel} \; \; \; \; ,
\eqno (4.4)
$$
$$
\Bigr[V_{M}(a_{41})-V_{M}(a_{311})\Bigr]>0
\Longleftrightarrow
12b+60(1-\gamma)>150 \mid \Omega \mid \;
{\parallel {\bf a} \parallel} \; \; \; \; ,
\eqno (4.5)
$$
$$
\Bigr[V_{M}(a_{41})-V_{M}(a_{2211})\Bigr]>0
\Longleftrightarrow
12b+60(1-\gamma)>-225 \mid \Omega \mid \;
{\parallel {\bf a} \parallel} \; \; \; \; ,
\eqno (4.6)
$$
$$
\Bigr[V_{M}(a_{311})-V_{M}(a_{2211})\Bigr]>0
\Longleftrightarrow
12b+60(1-\gamma)>-450 \mid \Omega \mid \;
{\parallel {\bf a} \parallel} \; \; \; \; .
\eqno (4.7)
$$
In light of Eqs. (4.2)-(4.7), if Eq. (4.5) is satisfied,
this ensures that all remaining conditions hold. One thus
obtains the inequality
$$
b>5(\gamma -1)+\left[{25\over 2} \mid \Omega \mid \;
{\parallel {\bf a} \parallel}
+O \Bigr({\parallel {\bf a} \parallel}^{2}\Bigr)
\right]=b_{0} \; \; \; \; ,
\eqno (4.8)
$$
which is a necessary and sufficient condition for the validity
of Eq. (4.1) when the Taylor expansion (3.9) is a good approximation.
Note that the term in square brackets on the r.h.s. of Eq. (4.8)
is a small correction of the value $5(\gamma -1)<0$ provided
${\parallel {\bf a} \parallel} \rightarrow 0$, as one would expect when
the Taylor expansion makes sense. Interestingly, the inequalities (4.1)
still hold for negative values of $b$ provided Eq. (4.8) is
satisfied, whereas the flat-space tree-level potential
${\hat V}_{M}={\hat b}\sum_{i=1}^{5}a_{i}^{4}$ used in [14] leads
to the value $b_{0}=0$.

Moreover, the reversed hierarchy (cf. (4.1))
$$
V_{M}(a_{321})>V_{M}(a_{2211})>V_{M}(a_{311})>V_{M}(a_{41})
\eqno (4.9)
$$
holds provided the following necessary and sufficient condition
is satisfied (cf. (4.4)) :
$$
b <5(\gamma -1)+\left[-{425\over 6}\mid \Omega \mid \;
{\parallel {\bf a} \parallel}
+O \Bigr({\parallel {\bf a} \parallel}^{2}\Bigr) \right]=b_{1}
\; \; \; \; .
\eqno (4.10)
$$
Again, the De Sitter background leads to a value $b_{1} \not =0$
with respect to the flat-space tree-level-potential result
$b_{1}=b_{0}=0$.

This preliminary analysis should be supplemented by a more
detailed numerical study. The aim of this investigation is
to prove that, for {\it all} values of ${\parallel {\bf a} \parallel}$
and $b$, the phase transition occurs only in the
$SU(3) \times SU(2) \times U(1)$ or
$SU(4) \times U(1)$ directions. From our previous discussion
(see also Figures 4-6), when $b \rightarrow + \infty$ the
absolute minimum is in the
$SU(3) \times SU(2) \times U(1)$ direction. However, if we
compute for fixed ${\parallel {\bf a} \parallel}$ the negative
${\overline b}^{0},{\overline b}^{1},{\overline b}^{2}$
values of $b$ such that
$$
V_{M}\left({\overline b}^{0},a_{321}\right)
=V_{M}\left({\overline b}^{0},a_{41}\right)
\; \; \; \; ,
\eqno (4.11)
$$
$$
V_{M}\left({\overline b}^{1},a_{321}\right)
=V_{M}\left({\overline b}^{1},a_{2211}\right)
\; \; \; \; ,
\eqno (4.12)
$$
$$
V_{M}\left({\overline b}^{2},a_{321}\right)
=V_{M}\left({\overline b}^{2},a_{311}\right)
\; \; \; \; ,
\eqno (4.13)
$$
we find ${\overline b}^{0}>{\overline b}^{1}$ and
${\overline b}^{0}>{\overline b}^{2}$,
$\forall \; {\parallel {\bf a} \parallel}$. This means that the
continuous transition to (4.9) leads to the interchanging of the
$SU(3) \times SU(2) \times U(1)$ with the
$SU(4) \times U(1)$ absolute minimum. Of course, similar
interchanges also occur for the relative minima, but they do not
affect the phase transition of the universe.

Defining
$$V_{M}^{(P)} \equiv \sum_{i=1}^{5} a_{i}^{4} \; \; \; \; ,
\eqno (4.14)
$$
and using Eqs. (3.13)-(3.16), it is useful to bear in mind the
formulae for ${\overline b}^{0}, {\overline b}^{1}$ and
${\overline b}^{2}$ obtained from Eqs. (4.11)-(4.13) :
$$
{\overline b}^{0}= {\Bigr[{\cal A}(a_{321})-{\cal A}(a_{41})\Bigr]\over
\Bigr[V_{M}^{(P)}(a_{41})-V_{M}^{(P)}(a_{321})\Bigr]} \; \; \; \; ,
\eqno (4.15)
$$
$$
{\overline b}^{1}= {\Bigr[{\cal A}(a_{321})-{\cal A}(a_{2211})\Bigr]\over
\Bigr[V_{M}^{(P)}(a_{2211})-V_{M}^{(P)}(a_{321})\Bigr]} \; \; \; \; ,
\eqno (4.16)
$$
$$
{\overline b}^{2}= {\Bigr[{\cal A}(a_{321})-{\cal A}(a_{311})\Bigr]\over
\Bigr[V_{M}^{(P)}(a_{311})-V_{M}^{(P)}(a_{321})\Bigr]} \; \; \; \; .
\eqno (4.17)
$$
The differences $\Bigr({\overline b}^{0}-{\overline b}^{1}\Bigr)$
and $\Bigr({\overline b}^{0}-{\overline b}^{2}\Bigr)$ are
plotted in Figures 4-6 as functions of
${\parallel {\bf a} \parallel}^{1\over 2}$
using Eqs. (4.15)-(4.17).
\vskip 1cm
\leftline {\bf 5. Concluding remarks}
\vskip 1cm
This paper has shown that the results in [14] about the
$SU(n)$ symmetry breaking in flat space may be generalized to a
curved, cosmological background such as De Sitter space.

The results in [13] have been thus re-obtained, by virtue of the
properties of the ${\cal A}$ function (Eq. (2.4) and Figures 1-3). They
confirm that the absolute minimum of the complete
one-loop potential
lies either in the $SU(3) \times SU(2) \times U(1)$ or in the
$SU(4) \times U(1)$ direction.
This provides a better understanding (cf. [13])
of the instability of the $SU(3) \times U(1) \times U(1) \times R_{311}$ and
$SU(2) \times SU(2) \times U(1) \times U(1) \times R_{2211}$ extrema,
since very simple and basic algebraic and numerical techniques have been used
(cf. Sect. 4).

Interestingly,
we can extend all our results to the most general and renormalizable
tree-level potential also containing cubic terms, since
the tree-level potential does not affect the
one-loop contribution within  the Coleman-Weinberg approach [12,13,15],
and the presence of an additional cubic term in $V_{M}^{(P)}$ (see (4.14))
favours the directions $a_{41}$ and $a_{321}$ (for which $V_{M}^{(P)}<1$)
with respect to $a_{311}$ and $a_{2211}$ (for which $V_{M}^{(P)}=0$).
The $SU(n)$ symmetry-breaking pattern for this more general class of
potentials in flat space can be found in [16], where the
author extends and confirms the results obtained in [14].
The approach considered above might be used to discuss the general
case of arbitrary directions in the adjoint representation of $SU(5)$;
one expects, however, that even in this more general case the absolute
minimum will be in the directions found by limiting the analysis to the
one-dimensional orbits.

The method here described may be applied to other GUT theories,
e.g. with $SO(10)$ or $E_{6}$ gauge groups, in De Sitter space [13].
These models appear as more realistic candidates for a unified theory
of non-gravitational interactions [8].
One would then obtain a physically more
relevant application of the techniques used in this paper.
\vskip 100cm
\leftline {\bf References}
\vskip 1cm
\item {[1]}
Weinberg S 1972 {\it Gravitation and Cosmology}
(New York: John Wiley and Sons)
\item {[2]}
Guth A H 1981 {\it Phys. Rev. D} {\bf 23} 347
\item {[3]}
Ellis G F R and Stoeger W 1988 {\it Class. Quantum Grav.} {\bf 5} 207
\item {[4]}
Padmanabhan T and Seshadri T R 1988 {\it Class. Quantum Grav.} {\bf 5} 221
\item {[5]}
Linde A D 1984 {\it Rep. Progr. Phys.} {\bf 47} 925
\item {[6]}
Hawking S W 1984 {\it Nucl. Phys. B} {\bf 239} 257
\item {[7]}
Esposito G and Platania G 1988 {\it Class. Quantum Grav.} {\bf 5} 937
\item {[8]}
O'Raifeartaigh L 1986 {\it Group Structure of Gauge Theories}
(Cambridge: Cambridge University Press)
\item {[9]}
Georgi H and Glashow S L 1974 {\it Phys. Rev. Lett.} {\bf 32} 438
\item {[10]}
Becker-Szendy R et al. 1990 {\it Phys. Rev. D} {\bf  42} 2974
\item {[11]}
Buccella F, Miele G, Rosa L, Santorelli P and Tuzi T 1989 {\it Phys. Lett. B}
{\bf 233} 178
\item {[12]}
Allen B 1983 {\it Nucl. Phys. B} {\bf 226} 228
\item {[13]}
Allen B 1985 {\it Ann. Phys.} {\bf 161} 152
\item {[14]}
Buccella F, Ruegg H and Savoy C A 1980 {\it Nucl. Phys. B} {\bf 169} 68
\item {[15]}
Coleman S and Weinberg E 1973 {\it Phys. Rev. D} {\bf 7} 1888
\item {[16]}
Ruegg H 1980 {\it Phys. Rev. D} {\bf 22} 2040
\vskip 100cm
\leftline {\bf Figure captions}
\vskip 1cm
\item {{\bf Figure 1.}} Differences of
${\cal A}$ values corresponding to (a)
full curve $\Bigr[{\cal A}(a_{41})
-{\cal A}(a_{321})\Bigr]$, (b) broken curve
$\Bigr[{\cal A}(a_{311})
-{\cal A}(a_{321})\Bigr]$ and (c) dotted curve
$\Bigr[{\cal A}(a_{2211})
-{\cal A}(a_{321})\Bigr]$. They are evaluated using
the Taylor expansion (3.9).

\item {{\bf Figure 2.}} Differences of logarithms$_{10}$ of ${\cal A}$
values corresponding to

{\item{(a)} full curve $\log_{10}\Bigr[
{\cal A}(a_{41})/{\cal A}(a_{321})\Bigr]$,
\item{(b)} broken curve $\log_{10}\Bigr[
{\cal A}(a_{311})/{\cal A}(a_{321})\Bigr]$ and
\item{(c)} dotted curve $\log_{10}\Bigr[
{\cal A}(a_{2211})/{\cal A}(a_{321})\Bigr]$.}

They are obtained using the exact formula (2.4) defining
${\cal A}(z)$.

\item {{\bf Figure 3.}} Differences of logarithms$_{10}$ of ${\cal A}$
values corresponding to

{\item{(a)} full curve $\log_{10}\Bigr[
{\cal A}(a_{41})/{\cal A}(a_{321})\Bigr]$,
\item{(b)} broken curve $\log_{10}\Bigr[
{\cal A}(a_{311})/{\cal A}(a_{321})\Bigr]$ and
\item{(c)} dotted curve $\log_{10}\Bigr[
{\cal A}(a_{2211})/{\cal A}(a_{321})\Bigr]$.}

The asymptotic expansion (3.10) is here applied.

\item {{\bf Figure 4.}} The dotted curve corresponds to the difference
$\Bigr({\overline b}^{0}-{\overline b}^{1}\Bigr)$, and the full curve
corresponds to the difference
$\Bigr({\overline b}^{0}-{\overline b}^{2}\Bigr)$,
where ${\overline b}^{0}$, ${\overline b}^{1}$ and ${\overline b}^{2}$
have been obtained in Eqs. (4.15)-(4.17). The Taylor expansion (3.9)
is here used for ${\cal A}(z)$.

\item {{\bf Figure 5.}} Dotted and full curve have the same meaning
as in Figure 4. The exact formula (2.4) is here used for ${\cal A}(z)$.

\item {{\bf Figure 6.}} Dotted and full
curve are defined as in Figures 4 and 5.
The asymptotic expansion (3.10) of ${\cal A}(z)$ is here applied.
\bye